\documentclass[useAMS,usenatbib]{mn2e}
\usepackage{url}
\usepackage{graphicx}
\font\japit = cmti10 at 10truept

\title
     [Supernova ejecta in SN1006]
{\vglue-3.0truecm
\centerline{\japit Submitted to Monthly Notices}
\vglue 2.5truecm
\noindent
A high resolution UV absorption spectrum of supernova ejecta in SN1006
\author[A. J. S. Hamilton et al.]
	{Andrew J. S. Hamilton$^1$, Robert A. Fesen$^2$, William P. Blair$^3$ \\
	$^1$JILA and Dept.\ Astrophysical \& Planetary Sciences,
	Box 440, U. Colorado, Boulder CO 80309, USA; Andrew.Hamilton@colorado.edu \\
	$^2$Dartmouth College, Hanover, NH 03755, USA; fesen@snr.dartmouth.edu \\
	$^3$Johns Hopkins University, Baltimore, MD 21218; wpb@pha.jhu.edu}
}

\newcommand{\dd}{{\rmn d}}	% MNRAS
\newcommand{\kms}{{\rm km} \ {\rm s}^{-1}}

\newcommand{\aap}[2]{A\&A, #1, #2}

\newcommand{\aj}[2]{AJ, #1, #2}

\newcommand{\apj}[2]{ApJ, #1, #2}
\newcommand{\apjs}[2]{ApJS, #1, #2}
\newcommand{\araa}[2]{ARA\&A, #1, #2}

\newcommand{\mn}[2]{MNRAS, #1, #2}
\newcommand{\nat}[2]{Nature, #1, #2}

\newcommand{\smrawfig}{
    \begin{figure*}
    \begin{minipage}{175mm}
    \includegraphics[scale=.98]{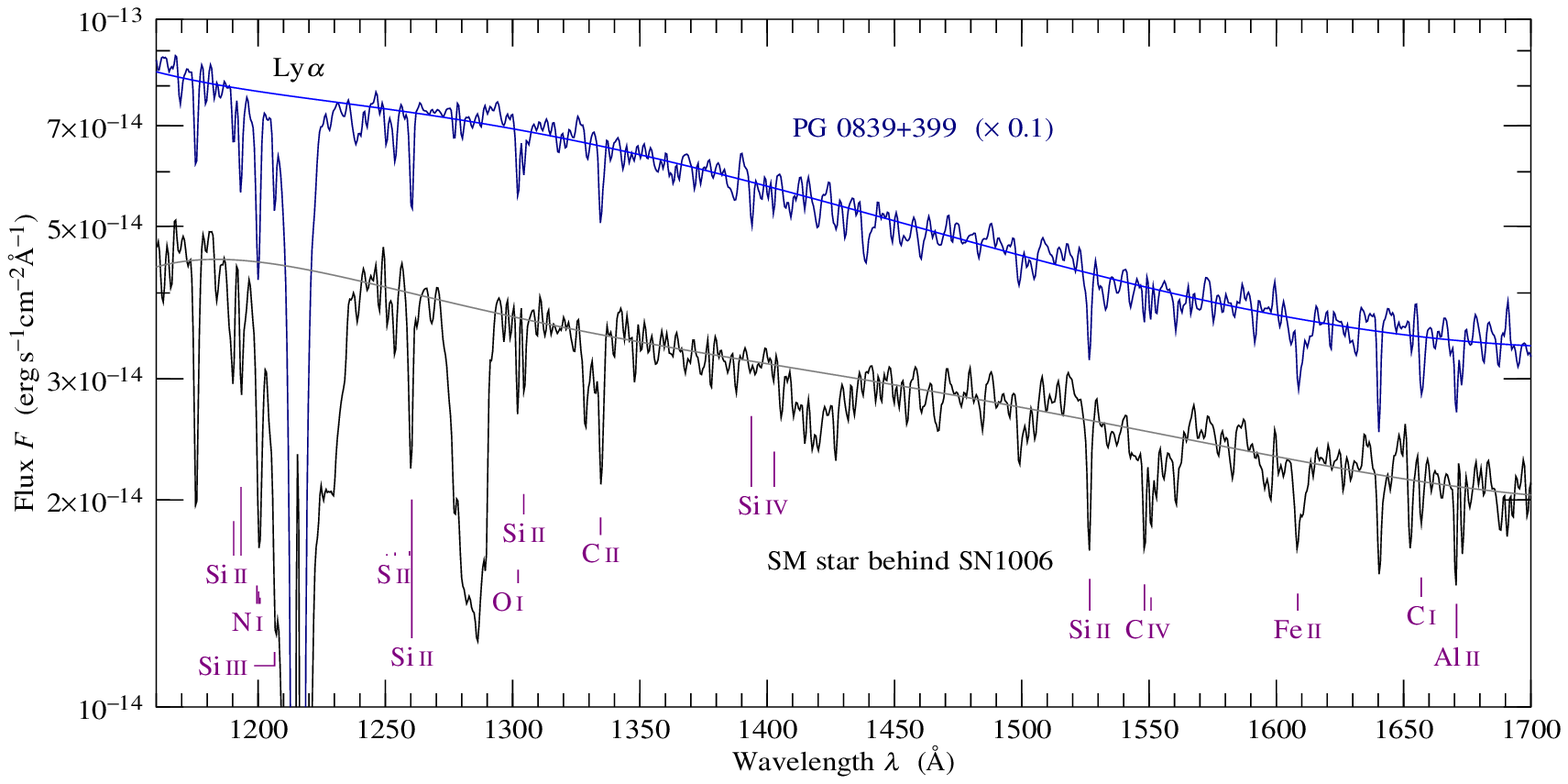}
    \caption[1]{
    \label{smraw}
STIS E140M spectra
of the SM star behind SN1006
and of the comparison star PG\ 0839+399
(the latter multiplied by a factor $0.1$ to bring it on to the graph).
Both spectra have been smoothed
to a resolution of $320 \  \kms$ FWHM.
The smooth lines are fitted continua,
quintic polynomials in $\log F$ as a function of $1/\lambda$.
Narrow interstellar lines are marked.
    }
    \end{minipage}
    \end{figure*}
}

\newcommand{\smfig}{
    \begin{figure*}
    \begin{minipage}{175mm}
    \includegraphics[scale=.98]{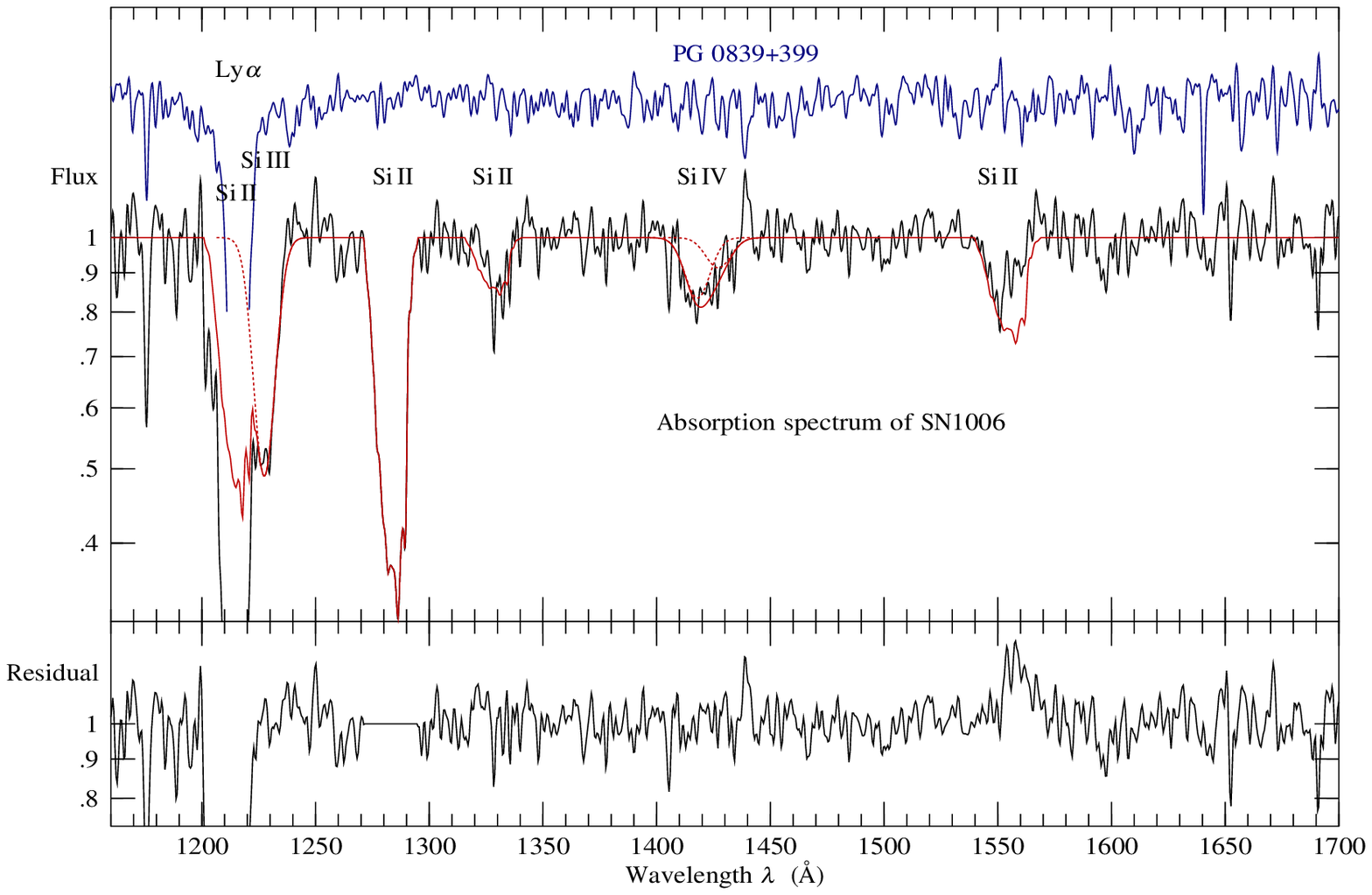}
    \caption[1]{
    \label{sm}
Absorption spectrum of the SM star,
showing the broad redshifted absorption features of Si~II, III, and IV
produced by supernova ejecta in SN1006.
The absorption spectrum is the ratio of
interstellar-line-excised,
smoothed,
continuum-corrected E140M spectra
of the SM star and of the comparison star PG\ 0839+399, the latter shown at top.
The smooth line passing through the absorption spectrum is
a model
in which the weaker Si~II 1304~\AA\ and Si~II 1527~\AA\ features
are assumed to have the same profile as the strong Si~II 1260~\AA\ feature,
which is assumed to be optically thin,
while the Si~III 1206~\AA\ and Si~IV 1394, 1403~\AA\ features
are assumed to have the same Gaussian profile as that of the shocked component
of Si~II 1260~\AA,
as described in \S\protect\ref{si34sec}.
The bottom panel shows the residual absorption spectrum of the SM star
after dividing by the model spectrum.
    }
    \end{minipage}
    \end{figure*}
}

\newcommand{\siahiresfig}{
    \begin{figure*}
    \begin{minipage}{175mm}
    \begin{center}
    \leavevmode
    \includegraphics[scale=1]{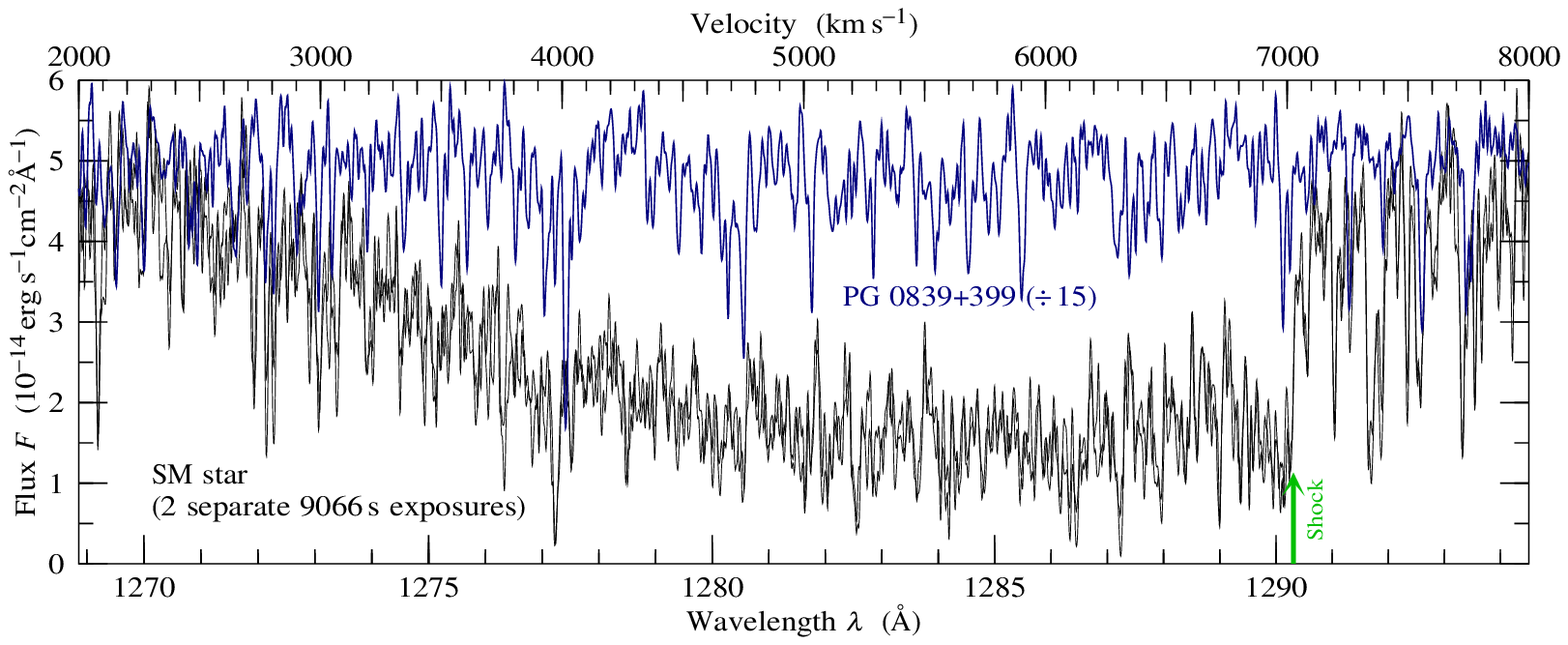}
    \end{center}
    \caption[1]{
    \label{si1260hires}
STIS E140M spectra 
of the redshifted Si~II 1260~\AA\ feature,
showing separately the two individual 9066 second exposures of the SM star,
and the comparison star PG\ 0839+399.
The spectrum of PG\ 0839+399 has been shifted by
$+ 32 \ \kms$ to mesh its stellar lines with those of the SM star,
and divided by $15$ to bring its continuum
to approximately same level as the SM star.
All spectra have been smoothed
with a near-Gaussian of FWHM
$6.8 \  \kms$.
The vertical arrow shows the position of the putative reverse shock
at $7026 \pm 10 \  \kms$.
The good agreement between
the two individual exposures of the SM star
demonstrates that the structure in the spectra is real, not noise.
Sadly,
the spectrum of the comparison star is not as similar in detail
to that of the SM star as one might have liked.
    }
    \end{minipage}
    \end{figure*}
}

\newcommand{\siafig}{
    \begin{figure}
    \begin{center}
    \leavevmode
    \includegraphics[scale=.6]{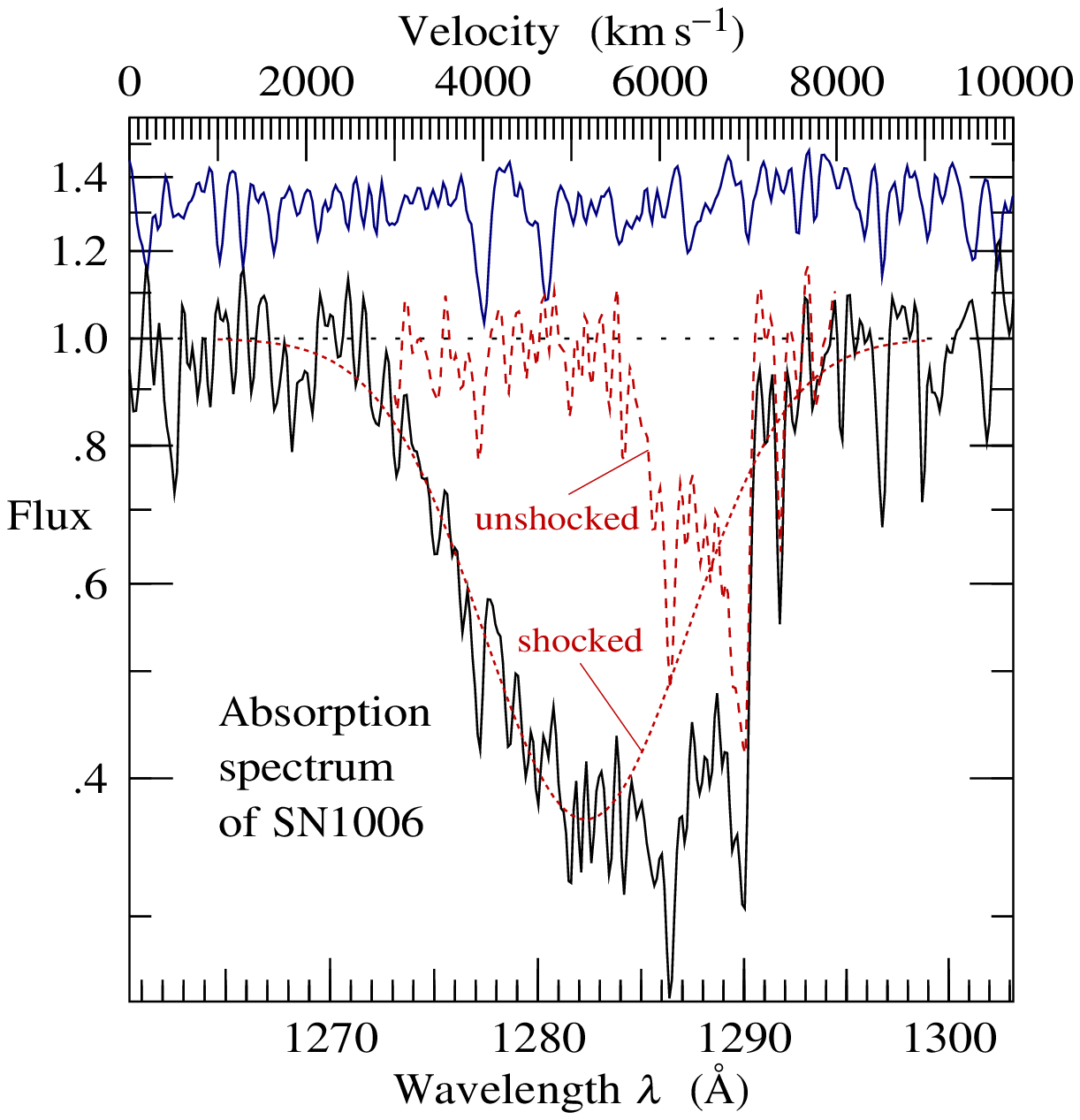}
    \end{center}
    \caption[1]{
    \label{si1260}
Absorption spectrum of SN1006 near the Si~II 1260~\AA\ feature,
showing the best fit Gaussian profile of shocked Si~II,
and the residual unshocked Si~II.
The spectrum is a higher-resolution zoomed-in version of the spectrum
shown in Figure~\protect\ref{sm}.
The spectrum is the ratio of
interstellar-line-excised,
continuum-corrected
spectra of the SM star
and of the comparision star PG\ 0839+399,
each smoothed to a resolution of $80 \  \kms$ FWHM
before their ratio was taken.
The upper spectrum shows the
interstellar-line-excised,
continuum-corrected comparison stellar spectrum
of PG\ 0839+399 at the same resolution,
shifted by
$+ 32 \ \kms$ to mesh its stellar lines with those of the SM star,
and offset to separate it from the SN1006 spectrum.
    }
    \end{figure}
}

\newcommand{\fosfig}{
    \begin{figure}
    \begin{center}
    \leavevmode
    \includegraphics[scale=.6]{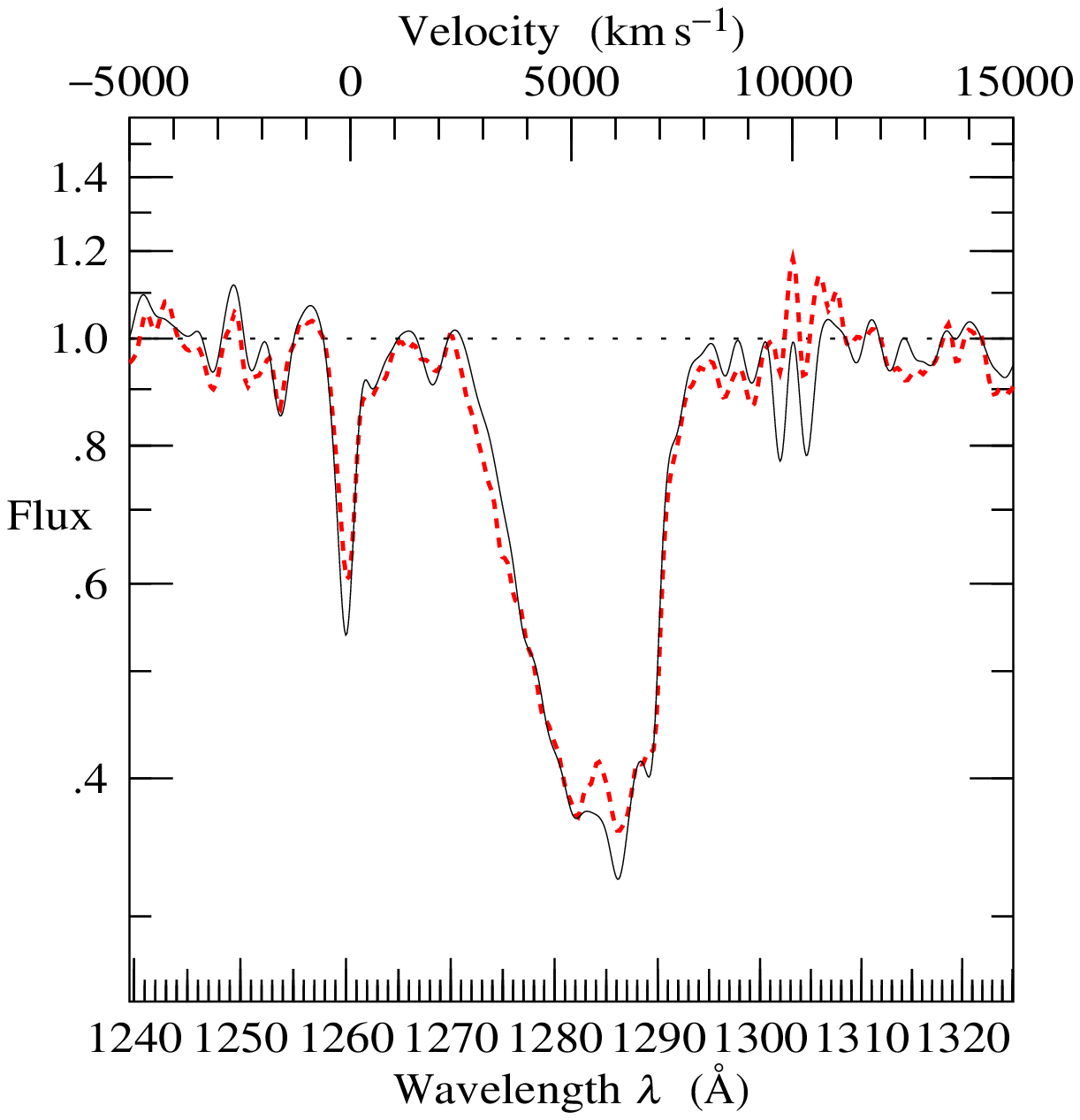}
    \end{center}
    \caption[1]{
    \label{fos}
Comparison of the FOS G130H (thick line)
to the STIS E140M (thin line)
absorption spectra of SN1006 near the Si~II 1260~\AA\ feature.
The FOS spectrum has been shifted by
$+ 41 \  \kms$
to bring its stellar and interstellar lines
into wavelength registration with the STIS spectrum,
and adjusted by a polynomial
to cancel a slowly varying difference in the FOS and STIS continua.
The STIS spectrum has been convolved to the resolution of the FOS spectrum,
$370 \  \kms$ FWHM.
Both spectra have been divided by
the smoothed comparison spectrum of PG\ 0839+399,
and corrected by the same quintic polynomial
applied to the ratio spectrum in Figure~\protect\ref{sm}.
    }
    \end{figure}
}

\newcommand{\partable}{
    \begin{table*}
    \begin{minipage}{175mm}
    \caption{Measured parameters}
    \label{par}
    \begin{tabular}{lrr}
Parameter
 & STIS (this paper)
 & FOS (Paper~1)
\\
Expansion velocity $r_s/t$ into reverse shock
 & $7026 \pm 3 \, (\mbox{relative})\pm 10 \, (\mbox{absolute}) \  \kms$
 & $7070 \pm 50 \  \kms$
\\
Mean velocity of shocked Si~II
 & $5160 \pm 70 \  \kms$
 & $5050 \pm 60 \  \kms$
\\
Dispersion $\sigma$ of shocked Si~II
 & $1160 \pm 50 \  \kms$
 & $1240 \pm 40 \  \kms$
\\
Reverse shock velocity $v_s$
 & $2680 \pm 120 \  \kms$
 & $2860 \pm 100 \  \kms$
\\
Shocked Si~II : Si~III : Si~IV
 & $1$ : $0.45 \pm 0.02$ : $0.32 \pm 0.02$
 & $1 \pm 0.03$ : $0.43 \pm 0.02$ : $0.41 \pm 0.02$
    \end{tabular}
    \end{minipage}
    \end{table*}
}

\newcommand{\obstable}{
    \begin{table*}
    \begin{minipage}{175mm}
    \caption{STIS Observing log}
    \label{obs}
    \begin{tabular}{lcccc}
Object & RA \& Dec (J2000) & Date & Orbits & Exposure (s) \\
PG\ 0839+399 & $08^{\rm h} 43^{\rm m} 12.70^{\rm s}$ $+39^\circ 44^\prime 49\farcs9$ & 21 Oct 1998 & 1 & 2{,}494 \\
SM star & $15^{\rm h} 02^{\rm m} 53.18^{\rm s}$ $-41^\circ 59^\prime 17\farcs6$ & 24-25 Jul 1999 & 8 & $2 \times ( 2{,}527 + 9{,}066 ) = 23{,}186$
    \end{tabular}
    \end{minipage}
    \end{table*}
}

\begin{document}

\maketitle

\begin{abstract}
We report a high resolution, far-ultraviolet,
STIS E140M spectrum of the strong, broad Si~II, III, and IV
features produced by the ejecta of SN1006
seen in absorption against the background Schweizer-Middleditch star.
%In an earlier paper,
%we suggested that the sharp edge on the Si~II 1260~\AA\ feature
%represented the location of the reverse shock moving into the ejecta.
The spectrum confirms the extreme sharpness
of the red edge of the redshifted Si~II 1260~\AA\ feature,
supporting the idea that this edge represents
the location of the reverse shock moving into the freely expanding ejecta.
The expansion velocity of ejecta at the reverse shock is measured to be
$7026 \pm 3 \, (\mbox{relative}) \pm 10 \, (\mbox{absolute}) \  \kms$.
If the shock model is correct,
then the expansion velocity should be decreasing at the observable rate of
$2.7 \pm 0.1 \  \kms$ per year.
The pre-shock velocity, post-shock velocity, and post-shock velocity
dispersion are all measured from the Si~II 1260~\AA\ feature,
and consistency of these velocities with the shock jump conditions
implies that there is little or no electron heating in this fast
($2680 \  \kms$) Si-rich shock.
\end{abstract}


%\clearpage

\section{Introduction}
\label{intro}

The supernova of 1006\,AD (SN1006) was the brightest in recorded history
\citep*{SG02}.
It is generally held that SN1006 was a probable Type~Ia,
the most straightforward evidence for this being its location
high
%($14.5^\circ$)
above the plane of the galaxy,
$\sim 500 \  {\rm pc}$ at the $2.18 \pm 0.08 \  {\rm kpc}$
distance measured by \citet*{WGL03}.
%far from any star-forming region.
The identification of SN1006 as Type~Ia is consistent with a broad
range of observational evidence and theoretical expectation.
The historical supernova was exceptionally bright,
and remained visible over an extended (3 year) period.
There is no sign of a pulsar or other central compact object.

The most likely cause of a Type~Ia supernova
is thought to be the thermonuclear explosion of a white dwarf
which accretes from a companion,
probably a giant or main sequence star,
but possibly a white dwarf
\citep{HN00,HGLM03,Hoflich06}.
Searches have so far failed to uncover
any candidate star that may have been the erstwhile companion in SN1006
\citep{RCSKMCFC03}.

\obstable

The remnant of SN1006 has been extensively observed
in the radio
\citep*{RG86,RG93,RMKWH88,MGR93},
optical
\citep*{KWC87,SKBW91,WL97,GWRL02,SGLS03,WGL03},
ultraviolet
\citep*{RBL95,KRZG04},
x-rays
\citep*{KPGHMOH95,WWPS96,WL97,VKBP00,APG01,DRBAP01,LRRWDP03,VLGRK03,BYUK03,DRB04,Vink05,KRBLCRR06}
and
$\gamma$-rays
\citep*{Tanimori98,TNY01,Aharonian05}.
The observations show
a limb-brightened, roughly circular blast wave expanding into
an interstellar medium of moderate density,
$\sim 0.1 \  {\rm cm}^{-3}$.
The x-ray image and spectrum reveal two components of emission:
limb-concentrated non-thermal synchrotron radiation
from relativistic electrons accelerated at the blast wave,
and more generally distributed lumpy thermal emission
from a hot plasma enhanced in heavy elements
\citep[][Fig.~2]{LRRWDP03}.
The thermal x-ray emission is consistent with what might be expected
from the ejecta of a Type~Ia supernova explosion
\citep[][Table~5]{DRBAP01},
but the inferences about masses and elemental abundances
are model dependent,
and alternative interpretations cannot be excluded
(but see \citealt{BBBD03} for a more optimistic view).

One of the more direct observational evidences on the nature of SN1006
comes from absorption spectroscopy
of a background OB subdwarf star discovered by
\citet[][hereafter the SM star]{SM80}
just $2.8^\prime$ south of the projected centre of the
$15^\prime$ radius remnant of SN1006.
Ultraviolet spectroscopy of the SM star with
the International Ultraviolet Explorer ({\sl IUE}) by
\citet*{WLSG83}
revealed broad absorption features,
which \citet{WLSG83} attributed to
Si~II, III, and IV, and Fe~II
in the ejecta of SN1006.
The presence of these features was confirmed in subsequent observations with
IUE \citep*{FH88},
and at higher resolution and signal-to-noise
with the Faint Object Spectrograph (FOS)
on the Hubble Space Telescope ({\sl HST})
\citep*{WCFHS93,WCHFLS97}.
\citet*{BLR96}
used the Hopkins Ultraviolet Telescope ({\sl HUT})
to observe the spectrum shortward of Ly$\alpha$,
finding marginal evidence for the presence of absorption by
Fe~III 1123~\AA.
Optically,
the SM star shows no evidence for absorption by SN1006 ejecta
\citep*{BHDB00}.
Recently,
\citet*{WLFH05}
have reported observations of four more UV sources behind SN1006,
finding clear evidence of absorption by SN1006 ejecta
in the two sources nearer to the projected centre.

In an extensive analysis of the {\sl HST\/} FOS spectrum of the SM star,
\citet[][hereafter Paper~1]{HFWCS97}
argued that the unusual shape of the broad Si~II 1260~\AA\ feature,
the strongest absorption feature in the 1150--2000~\AA\ ultraviolet spectrum
of the SM star,
arises because the feature consists of both unshocked and shocked components.
This surprising conclusion was driven by several observational facts.
(1) The Si~II 1260~\AA\ feature had a sharp red edge,
at $7070 \pm 50 \  \kms$,
which Paper~1 identified as the position of the reverse shock front
in the freely expanding ejecta of SN1006.
This suggested that the reverse shock in SN1006
is currently moving into ejecta containing Si~II.
%This suggested that unshocked Si~II is present.
%and is undergoing a shock.
(2) The depth of absorption of the Si~II 1260~\AA\ feature
implied that the collisional ionization time of shocked Si~II
is of the order of the 1000~yr age of the remnant.
This implied that there should also be Si~II in the shocked ejecta.
%This implied that shocked Si~II should also be present.
%if unshocked Si~II is present and undergoing a shock.
(3) The blue edge of the Si~II 1260~\AA\ feature fitted nicely to a
broad Gaussian profile, as expected for shocked Si~II,
and the same Gaussian profile accounted naturally for the slight tail
of absorption observed to high velocities
$> 7000 \  \kms$.
(4) The velocities of unshocked and shocked gas
measured from the observed Si~II 1260~\AA\ line profile
were consistent with the shock jump conditions.

In this paper we report
Hubble Space Telescope Imaging Spectrometer (STIS)
observations of the SM star behind SN1006,
with a spectral resolution of
$5 \  \kms$ FWHM,
$\sim 60$ times the resolution of the earlier FOS spectrum
\citep{WCHFLS97}.
A primary motivation for obtaining the high resolution STIS spectrum
was to test the proposal of Paper~1
that the Si~II 1260~\AA\ feature
indeed reveals ejecta in SN1006 undergoing a reverse shock.
A key prediction of the shock model is that
%High resolution spectra can test two predictions of the shock model.
%First,
the red edge of the Si~II 1260~\AA\ feature,
representing the position of the reverse shock,
should be extremely sharp,
since a shock should decelerate gas `instantaneously'
from the free expansion velocity to some lower bulk velocity.
High resolution spectroscopy can also test for possible problems,
such as saturation, or contamination by narrow lines.

%Although the stellar spectrum of the SM star is relatively featureless,
%it is not entirely so,
One of the main sources of unquantified systematic uncertainty
in interpreting the absorption features in the FOS spectrum
was the possibility of contamination by stellar lines.
We therefore obtained and report here STIS observations of
a comparison star PG\ 0839+399
\citep*{GSL86}
identified by
\citet{BLR96}
as having not only a similar temperature and gravity,
but also similar photospheric abundances,
and low extinction.

The STIS spectra reported here do not cover the
Fe~II 2383~\AA\ and 2600~\AA\ absorption features
observed in the SM star with IUE
\citep{WLSG83,FH88}
and the FOS
\citep{WCFHS93,WCHFLS97},
and more recently
in a second background source, the quasar QSO-1503-4155,
with low-resolution STIS observations by
\citet{WLFH05}.

\section{Observations}
\label{observations}

\smrawfig

\smfig

\partable

\siahiresfig

Far-ultraviolet echelle spectra of
the SM star and of the comparison star PG\ 0839+399
were obtained
in 1998 and 1999
with STIS
using
the $2^{\prime\prime} \times 2^{\prime\prime}$ slit,
the E140M grating,
and the FUV-MAMA detector.
The observing log is given in Table~\ref{obs}.
The spectra cover 1150--1700~{\AA}
in wavelength bins
$\Delta \lambda \approx \lambda / 92000$,
or $3.3 \  \kms$, wide.
Over the relevant range of the spectra from 1220--1540~\AA,
the spectral resolution is 1.4\ pixels, or $4.6 \  \kms$, FWHM.
The line spread function is non-Gaussian,
with a narrow core and broader wings.
%The spectral resolution in this configuration is
%$R \equiv \lambda / ( 2 \Delta \lambda ) = 45{,}800$,
%some $35$ times higher than the resolution
%$R = 1{,}300$ of the FOS observations reported by \citet{WCFHS93}.
Over 1220--1540~\AA,
the signal-to-noise ratio is $\sim 8$ per pixel and $\sim 15$ per pixel
for the SM star and PG\ 0839+399 respectively.

%The data were reduced using a standard package, Rob Fesen,
%who personally inserted fluxes at each wavelength for each object,
%thus allowing the original observations to be discarded,
%for a considerable saving in effort,
%consistent with current NASA policy.
%Even so, some `bad' pixels failed to meet expectations,
%and had to be forced to conform.
%Finally,
%a careful record of the entire reduction procedure was saved
%inside a black hole.
%All of this took several years,
%accounting for the unconscionable delay in publication of this
%eagerly anticipated tract.

Figure~\ref{smraw}
shows the STIS E140M spectra of the SM star
and of the comparison star PG\ 0839+399.
Each spectrum has been
convolved with a near-Gaussian
%(a function proportional to
%$\left\{ 1 - \Delta \lambda / [ (2n{+}3) \langle \Delta \lambda^2 \rangle ]^{1/2} \right\}^n$
%with $n = 4$)
with a FWHM of
$320 \  \kms$.
The near-Gaussian is a function proportional to
$( 1 - x^2 )^n$ for $| x | < 1$,
with
%$x \equiv$ $\Delta v / [ (2n{+}3) \langle \Delta v^2 \rangle ]^{1/2}$ and
$n = 4$.
Such a smoothing function
mostly shares the nice theoretical properties of Gaussian smoothing,
but extends over only a finite number of pixels.
The smoothing is broad so as to bring out
the broad absorption features in the spectrum of the SM star;
higher resolution versions of the spectra are
shown below in \S\ref{results}.
There are a few narrow gaps in the spectra longward of
$1600 \  {\rm \AA}$,
and the spectra were linearly interpolated across these
before being smoothed.

The small scale structure in the various spectra
in Figure~\ref{smraw}
is not noise,
but real, mostly stellar features.
This fact can be inferred from the good agreement
between spectra from the individual exposures of the SM star,
apparent for example in the high-resolution zoom of the Si~II 1260~\AA\ feature
shown in Figure~\ref{si1260hires} below.

The stellar lines of the comparison star PG\ 0839+399 are blueshifted,
by $- 32 \  \kms$,
relative to those of the SM star.
Narrow interstellar absorption lines
in PG\ 0839+399
show no discernible shift relative to those in the SM star.

Figure~\ref{sm}
shows the ratio of the spectrum of the SM star
to that of the comparison star PG\ 0839+399.
Before their ratio was taken,
both spectra were
linearly interpolated across narrow interstellar resonance lines,
the spectrum of PG\ 0839+399 was
redshifted
by $+ 32 \  \kms$
to mesh its stellar lines with those of the SM star,
and both spectra were smoothed to a resolution of $320 \  \kms$ FWHM,
the same as in Figure~\ref{smraw}.
The continuum was then removed by fitting
unabsorbed parts of the ratio spectrum
to a quintic polynomial in log flux
%$\log F$
versus inverse wavelength.
%$1/\lambda$.
The continuum fitting is intended to remove
differences both in the intrinsic continua of the spectra
and in the extinction.
We tried correcting the continua for the extinction alone,
measured by \citet[][Table~2]{BLR96}
to be $E_{B-V} = 0.119 \pm 0.005$ for the SM star
and $E_{B-V} = 0.045 \pm 0.005$ for PG\ 0839+399,
using the extinction curve of \citet*{CCM89},
but this correction leaves noticeable differences in the continua,
particularly at shorter wavelengths,
so we preferred the empirical polynomial fit.

The ratio spectrum of Figure~\ref{sm}
shows broad absorption features,
which, following \citet{WLSG83}
and subsequent authors
\citep{FH88,WCFHS93,WCHFLS97,HFWCS97},
we attribute to
redshifted Si~II, III, and IV in the ejecta of SN1006.

As in the individual spectra shown in Figure~\ref{smraw},
the small scale structure in the ratio spectrum of Figure~\ref{sm}
is not noise, but real, mostly stellar features.
It is thus apparent that the spectral type and elemental abundances
of the comparison star PG\ 0839+399
are not identical to those of the SM star.

The smooth line passing through the absorption spectrum
in Figure~\ref{sm} is a model
in which the weaker Si~II 1304~\AA\ and Si~II 1527~\AA\ features
are assumed to have the same profile as the strong Si~II 1260~\AA\ feature,
with strengths in proportion to their wavelengths times oscillator strengths,
as expected if the lines are optically thin.
The Si~III 1206~\AA\ and Si~IV 1394, 1403~\AA\ features
are assumed to have Gaussian profiles with the same width as
that measured in \S\ref{si1260sec}
for the shocked component of the Si~II 1260~\AA\ feature.

\section{Results}
\label{results}

\subsection{Si~II 1260~\AA\ feature}
\label{si1260sec}

%\siarawfig

Figure~\ref{si1260hires}
shows the STIS spectrum of the SM star at high resolution
in the vicinity of the redshifted Si~II 1260~\AA\ absorption feature.
The Figure shows individually the two long exposure spectra,
$9066 \  {\rm s}$ each, of the SM star,
and the spectrum of the comparison star PG\ 0839+399,
the last shifted by $+ 32 \  \kms$ to mesh its stellar lines with those of the SM star,
and divided by $15$ to bring its continuum to approximately the
same level as that of the SM star.
Each spectrum has been smoothed with a near Gaussian
of $6.8 \  \kms$ FWHM,
just slightly greater than the
$4.6 \  \kms$
intrinsic resolution of the spectrum.

The arrow
at $7026 \  \kms$
marks the velocity of the sharp red edge of the feature,
the putative position of the reverse shock
moving into the freely expanding Si~II ejecta.

The spectrum would seem to confirm the key prediction of the shock hypothesis,
that the edge of the feature should be extremely sharp.
At the steepest point,
the flux shown in Figure~\ref{si1260hires}
rises over an interval of 7 pixels
from an absorption fraction of
$\sim 0.3$ at $7016 \  \kms$
to an absorption fraction of
$\sim 0.8$ at $7036 \  \kms$.
%whereafter the flux continues to rise somewhat less steeply,
%to a maximum near
%$7050 \  \kms$.
We take the position of the reverse shock to be
$7026 \  \kms$,
at the centre of the steep interval,
and this is the value given in Table~\ref{par}.

We have compared the shape of the steep interval to that expected
for a step function smoothed over the line spread function of STIS
with the E140M grating and $2^{\prime\prime} \times 2^{\prime\prime}$ slit,
and conclude that the edge is consistent with being unresolved.
%or barely resolved.

It is natural to assign the uncertainty of the location of the
reverse shock to be $\pm 10 \  \kms$,
since that is the extent of the steep interval.
However, if at some time in the future
a second epoch observation is taken at similar resolution and signal-to-noise,
then by comparing the line profiles near the shock,
it should be possible to measure the change in the position
of the shock with significantly higher resolution.
Given the quality of the present spectrum,
we judge that it should be possible to measure the change in the position
of the shock to better than a pixel, about $3 \  \kms$.
We therefore choose to quote
in Table~\ref{par}
two separate errors on the
$7026 \  \kms$ position of the shock,
a relative eror of
$\pm 3 \  \kms$,
which is an estimate of the uncertainty in the relative position
of the shock between two observations of similar quality
at two different times,
and an absolute error of
$\pm 10 \  \kms$,
which is an estimate of the uncertainty in the absolute position
of the shock front in the freely expanding ejecta.

\siafig

Figure~\ref{si1260hires}
also supports a second proposition of the shock model,
that the Si~II 1260~\AA\ absorption feature is optically thin.
If the feature were produced by dense, optically thick knots of ejecta,
a possibility discussed by \citet{FWLH88} and \citet{FH88},
then the absorption feature would be choppier,
hitting zero at various wavelengths.
The deepest part of the absorption trough,
over 1280--1290~\AA,
remains clear above zero.

Figure~\ref{si1260}
shows the absorption spectrum of SN1006,
the ratio of the
spectrum of the SM star to that of the comparison star PG\ 0839+399,
in the vicinity of the Si~II 1260~\AA\ absorption feature.
The Figure is essentially a zoomed-in version of Figure~\ref{sm},
but at four times the spectral resolution.
As with Figure~\ref{sm},
both spectra were first
linearly interpolated across narrow interstellar resonance lines,
and the spectrum of PG\ 0839+399 was
redshifted by $+ 32 \  \kms$.
Both spectra were smoothed with a near Gaussian of
$80 \  \kms$ FWHM
before their ratio was taken,
and the ratio was then corrected by the same quintic polynomial
as in Figure~\ref{sm}.

The maximum velocity $7026 \  \kms$ of unshocked Si~II
represents the position of the reverse shock,
and as discussed in Paper~1,
Si~II passing through the reverse shock will be decelerated suddenly
from the free expansion velocity of $7026 \  \kms$
to a lower bulk velocity.
The collisional ionization time of the shocked Si~II
is comparable to the age of the remnant,
so it is expected that Si~II should persist after being shocked.
The shocked Si~II should contribute a broad Gaussian profile of absorption.
Fitting the blue edge of the Si~II 1260~\AA\ feature to a Gaussian,
we find the best fit shown in Figure~\ref{si1260}.
The best fit parameters of the Gaussian fit
are a mean of
$5160 \pm 70 \  \kms$,
and a dispersion of
$1160 \pm 50 \  \kms$,
as listed in Table~\ref{par}.

Table~\ref{par} also gives, for reference,
values previously measured from the FOS data in Paper~1.
The FOS velocities listed in Table~\ref{par}
are as given by Paper~1,
and have not been corrected for the $- 41 \  \kms$
shift of the FOS spectrum
measured in \S\ref{evolution} below.
It is to be noted that the uncertainties on the STIS parameters are,
except for the much more precise measurement of
the expansion velocity into the reverse shock,
not much different from the quoted uncertainties on the FOS parameters.
As in Paper~1,
the stated uncertainties are formal 1 standard deviation statistical errors,
and do not include the systematic uncertainty associated
with the stellar spectrum or with placement of the continuum.
In the present paper the systematics are under better control,
thanks to the template stellar spectrum from the comparison star PG\ 0839+399,
so that, notwithstanding the comparable statistical errors,
the STIS values in Table~\ref{par} are more reliable, and are to be preferred.

The consistency of the measured velocities can be checked,
as in Paper~1,
against the jump conditions for a strong shock.
The shock jump conditions predict that
the three-dimensional velocity dispersion $3^{1/2} \sigma$ of the ions
should be related to
the deceleration $\Delta v$ of the shocked gas
by energy conservation
\begin{equation}
\label{check}
  3^{1/2} \sigma
  =
  \Delta v
\end{equation}
provided that all the shock energy goes into ions.
From the measured 1-dimensional dispersion
$\sigma = 1160 \  \kms$
given in Table~\ref{par},
the observed 3-dimensional dispersion is
\begin{equation}
  3^{1/2} \sigma
  =
  3^{1/2}
  \times
  ( 1160 \pm 50 \  \kms )
  =
  2010 \pm 90 \  \kms
  \ .
\end{equation}
The observed deceleration is
the difference between the measured free expansion velocity
$7026 \  \kms$ of the unshocked gas at the reverse shock front,
and the measured mean velocity $5160 \  \kms$ of the Gaussian velocity distribution of shocked gas,
which is
\begin{equation}
  \Delta v
  =
  ( 7026 \pm 10 ) - (5160 \pm 70)
  =
  1870 \pm 70 \  \kms
  \ .
\end{equation}
As previously found in Paper~1,
these values are in good agreement
(the observed dispersion exceeds that predicted from the deceleration
by $1.2$ standard deviations),
encouraging the view that the shock interpretation is basically correct.

The reverse shock velocity $v_s$ corresponding to the observed dispersion
$\sigma$ is
\begin{equation}
\label{vs}
  v_s = (16/3)^{1/2} \sigma = 2680 \pm 120 \  \kms
\end{equation}
which is listed in Table~\ref{par}.

As remarked above,
the relation~(\ref{check})
is valid only if all the shock energy goes into ions,
with negligible energy going into heating electrons.
If some of the shock energy went into electrons,
then the predicted ion velocity dispersion $\sigma$ would be decreased,
which would worsen the agreement with the shock jump conditions.
Thus, if the shock interpretation of the Si~II 1260~\AA\ feature is accepted,
then the measured velocities constitute
direct evidence that little electron heating occurs in shocks of this kind.
This is consistent with other observational evidence
that little electron heating occurs in the fast shocks in SN1006
\citep{LRMB96,GWRL02,VLGRK03,Vink05}.

\subsection{Evolution of the reverse shock}
\label{evolution}

\fosfig

As the reverse shock propagates into the Si~II ejecta,
the free expansion velocity $r_s / t$
at the position $r_s$ of the reverse shock should decrease with time $t$ as
\begin{equation}
\label{drs}
  {\dd ( r_s / t ) \over \dd t}
  =
  - {v_s \over t}
  \ .
\end{equation}
Given the measured shock velocity $v_s$, equation~(\ref{vs}),
and the known $t = 993 \  {\rm yr}$
age of the remnant at the 1999 epoch of observation,
equation~(\ref{drs}) predicts that the expansion velocity
of the reverse shock should be decreasing at a rate of
\begin{equation}
  {\dd ( r_s / t ) \over \dd t}
  =
  - 2.7 \pm 0.1 \  \kms \  {\rm yr}^{-1}
  \ .
\end{equation}

We have attempted to measure the expected change in the position of
the reverse shock between the FOS and STIS data.
The time that elapsed between the FOS observations,
taken during 17--20 September 1993
\citep{WCHFLS97},
and the STIS observations, taken during 24--25 July 1999,
is 2135~days, or 5.85~years,
so if the shock model is correct
then it is expected that the position of the reverse shock should
have changed by
$\Delta ( r_s / t) = 5.85 \  {\rm yr} \times ( - 2.7 \pm 0.1 \  \kms {\rm yr}^{-1} )$,
that is by
\begin{equation}
\label{drspred}
  \Delta ( r_s / t) =
  - 15.8 \pm 0.7 \  \kms
\end{equation}
between the FOS and STIS observations.

To measure the change in the position of the reverse shock
as accurately as possible,
we carried out a differential comparison of the FOS
and STIS spectra.
The FOS spectrum was taken by
\citet{WCHFLS97}
using the
$1^{\prime\prime}$ diameter circular aperture,
the G130H grating,
and the blue detector.
%in which configuration the line spread function
%is approximately Gaussian with a dispersion of 
%$90 \  \kms$ at 1290~\AA.
The uncertainties in the comparison are dominated
by uncertainties in the FOS spectrum.
We took the error in each FOS spectral bin
from the photon-counting noise tabulated in the {\sl HST\/} spectral data file,
multiplied by an empirical factor of 1.2
suggested by the scatter in the flux difference between adjacent spectral bins.
Values and $1 \sigma$ uncertainties of parameters given below
are deduced from the minimum and deviation in the $\chi^2$
of the difference between the FOS and STIS spectra.

To compare the FOS and STIS spectra,
we first smoothed the STIS spectrum
with a near-Gaussian
to the resolution of the FOS spectrum.
Fitting the smoothed STIS spectrum to the FOS spectrum
yields a best fit resolution of
$370 \pm 10 \  \kms$ FWHM,
somewhat larger than the nominal
% FWHM = (1.27 ± .04 diodes) * (.305 ± .004 arcsec/diode) / (.0774 ± .001 arcsec/pixel) = 5.0045220 pixels
% 1 pixel is 58.657610 km/s at 1290 Å
$295 \pm 10 \  \kms$ FWHM of the line spread function of the FOS
G130H spectrum in the configuration used.
Comparison of stellar and interstellar lines
in unabsorbed (by SN1006 ejecta) ranges of the spectrum over
1240--1400\ {\AA}
indicate that the FOS spectrum is shifted by
$- 41 \pm 6 \  \kms$
relative to the STIS spectrum.
To bring the FOS lines into registration with those of STIS,
we therefore shifted the FOS spectrum by
$+ 41 \  \kms$.
The FOS and STIS continua differ somewhat,
the FOS continuum varying from $\sim 10\%$ high near 1170\ {\AA}
to about $\sim 10\%$ low near 1400\ {\AA}.
To cancel this variation,
we fitted the log of the FOS to STIS flux ratio
to a 19th order polynomial as a function of wavelength,
and divided the FOS spectrum by the fitted ratio.
Despite the high order,
the correcting polynomial varies moderately and smoothly with wavelength.

Figure~\ref{fos}
compares the resulting adjusted FOS spectrum
to the smoothed STIS spectrum
in the vicinity of the Si~II 1260~\AA\ feature.

Having made these preliminary adjustments,
we measured the change in the velocity of the shock front
between the FOS and STIS spectra
from the shift in wavelength of the sharp edge at 1289.5--1291~\AA.
The measured shift of the edge of the STIS spectrum
relative to the FOS spectrum is
\begin{equation}
\label{drsobs}
  \Delta ( r_s / t) =
  - 35 \pm 25 \  \kms
\end{equation}
in which approximately half the error
comes from uncertainty in the wavelength of the edge,
and the other half from an estimated $2\%$
uncertainty in the relative normalization of the
FOS and STIS continua.
We note that each FOS spectral bin is
$60 \  \kms$ wide,
so the error in the shift is about half an FOS spectral bin.

The measured change
$- 35 \pm 25 \  \kms$,
equation~(\ref{drsobs}),
in the position of the shock front
between the STIS and FOS data
is consistent with the predicted change
$- 15.8 \pm 0.7 \  \kms$,
equation~(\ref{drspred}).
However,
the difference is also only $1.4$ standard deviations from
no change at all,
and it is not yet possible to claim a detection.

\subsection{Si~III and Si~IV features}
\label{si34sec}

As in Paper~1,
we have fitted the Si~III 1206~\AA\ and Si~IV 1394, 1403~\AA\ 
features in the absorption spectrum of SN1006
to a combination of unshocked and shocked components,
the two components being assumed to have the same profiles
as the unshocked and shocked components of the Si~II 1260~\AA\ feature
shown in Figure~\ref{si1260}.
The results are essentially the same as found in Paper~1:
the Si~III and Si~IV profiles are consistent with being entirely shocked,
with no evidence of any unshocked component.
%The absence of unshocked Si~IV accounts for the difference in the Si~II
%and Si~IV profiles remarked by \citet{WCHFLS97}.

Figure~\ref{sm}
shows, in dotted lines,
the fitted Gaussian profiles of shocked Si~III and Si~IV,
the centre and width of the Gaussian in each case being fixed equal
to those measured from the Si~II 1260~\AA\ feature,
as given in Table~\ref{par}.
The best fit ratios of number densities of shocked Si~III and Si~IV
relative to shocked Si~II are listed in Table~\ref{par}.
As elsewhere in this paper and in Paper~1,
the quoted uncertainties are formal $1\sigma$ errors,
and do not include systematic uncertainties associated
with the stellar spectrum or with placement of the continuum.
The systematic uncertainty could well exceed the statistical uncertainty.
%especially in the case of the Si~IV abundance.

If the width of the Gaussian is permitted to be a free parameter,
the centre being held fixed,
then the Si~III profile prefers a slightly narrower profile,
$1070 \pm 40 \  \kms$,
while the Si~IV profile prefers a slightly broader profile,
$1220 \pm 140 \  \kms$,
but in both cases the width is consistent with the
$1160 \pm 50 \  \kms$
dispersion of the Si~II profile.

If not only the width but also
the centre of the Gaussian are permitted to be free parameters
(which yields a reliable result only for Si~IV,
since the blue edge of Si~III is confused by Ly$\alpha$),
then the Si~IV profile prefers a slightly broader width,
$1280 \pm 140 \  \kms$,
still consistent with that of Si~II,
but a lower central velocity of
$4850 \pm 100 \  \kms$,
which is 2--$3 \sigma$ below the
$5160 \pm 70 \  \kms$
centre measured from Si~II,
suggesting weakly that Si~IV may perhaps extend
to somewhat lower velocities than Si~II.
Examination of the Si~IV line profile, Figure~\ref{sm},
indicates that the fit is pulled to lower velocities
mainly by an absorption spike at 1405~\AA\ (not at Si~IV 1403~\AA!).
The spike is not so prominent in the FOS spectrum,
which does however show, like the STIS spectrum,
a slight excess of Si~IV absorption at lower velocities.

In Paper~1,
the Si~IV preferred a significantly broader width of
$1700 \pm 100 \  \kms$.
The narrower width measured here
is to a large extent a consequence of dividing by
the template stellar spectrum of PG\ 0839+399
--
compare the raw Si~IV profile of the SM star in Figure~\ref{smraw}
to the corrected Si~IV profile in Figure~\ref{sm}.

Since the collisional ionization time of Si~II is comparable to
the age of the remnant (Paper~1),
it is not surprising that collisional ionization of the shocked Si~II
should lead to observable amounts of shocked Si~III and Si~IV.

Irradiation of unshocked ejecta by UV and x-ray emission
from the reverse shock could in principle photoionize Si~II
to Si~III and Si~IV.
Indeed, Paper~1 remarked that
the ratios of the observed column densities of Si~II, III, and IV
are close to the unshocked ratios predicted
in photoionization trials similar to those described by \citet{HS84},
in which the deflagrated white dwarf model CDTG7
of Woosley (Woosley 1987, private communication),
similar to model CDTG5 of \citet{WW87},
is evolved into a uniform ambient medium.
Thus the fact that the observations point to negligible amounts
of unshocked Si~III and Si~IV
places constraints on the explosion model
-- for example, the model considered by \citet{HS84}
produces more unshocked Si~III and Si~IV than observed.

In Paper~1
it was argued that,
under a `simplest' set of assumptions, there should be an observable
column density of blueshifted Si~IV,
although this prediction is sensitive to the assumptions,
and is not robust.
The STIS spectrum, Figure~\ref{sm},
shows no evidence of blueshifted Si~IV absorption along this sight line,
in agreement with previous observations.

\subsection{Weaker Si~II features}

The absorption spectrum of SN1006 in Figure~\ref{sm} shows,
in addition to the broad, redshifted Si~II 1260~\AA\ feature,
weaker features of Si~II 1304~\AA\ and Si~II 1527~\AA.

\citet{FWLH88}
first pointed out that the Si~II 1527~\AA\ feature
is anomalously weak compared to the
principal Si~II 1260~\AA\ feature.
The discrepancy was confirmed by the FOS observations \citep{WCHFLS97},
and is confirmed again in the present observations,
Figure~\ref{sm}.

The high quality of the STIS spectra
establishes beyond doubt that the discrepancy is real,
not just noise.
As remarked in \S\ref{observations},
the small scale structure in the absorption spectrum shown in Figure~\ref{sm}
is real, not noise, and must be attributed,
at least in unabsorbed (by SN1006) regions of the spectra,
to differences between the stellar spectra of the SM star
and the comparison PG\ 0839+399 star.

The bottom panel of Figure~\ref{sm}
shows the residual spectrum obtained by dividing the
absorption spectrum by a model in which
the weaker Si~II 1304~\AA\ and Si~II 1527~\AA\ features
are assumed to have the same profile as the
strong Si~II 1260~\AA\ feature.
It is evident that the most prominent residual
(besides Ly$\alpha$)
is around the anomalous Si~II 1527~\AA\ feature.
By contrast,
the Si~II 1304~\AA\ feature appears to be consistent with Si~II 1260~{\AA},
in the sense that the level of fluctuation in the residual
is comparable to that in unabsorbed (by SN1006) parts of the spectrum.
The principal anomaly of the Si~II 1527~\AA\ feature
is that it is missing absorption on the red side of the line,
by perhaps a factor of 2.
The Si~II 1304~\AA\ feature does not share this anomaly.

Note that
the red side of the Si~II 1304~\AA\ feature is affected by interstellar
C~II 1334.5~\AA,
while the blue side of the Si~II 1527~\AA\ feature is affected by interstellar
C~IV 1548.2, 1550.8~\AA.
However,
these interstellar lines absorb only a narrow interval of the high
resolution STIS spectra, and their presence does not change any conclusions.

\citet{FWLH88} proposed that
the Si~II 1260~\AA\ feature might be augmented by
a contribution from
S~II 1250.6, 1253.8, 1259.5~\AA.
Paper~1 argued against this proposal on the grounds that
that the combined oscillator strength of the S~II multiplet,
$0.005 + 0.011 + 0.016 = 0.033$,
is only 1/30th that of the Si II 1260.4~\AA\ line,
whose oscillator strength is $1.007$
\citep*{Morton91}.
This, along with the fact that explosion models
predict that S and Si are synthesized together
in a cosmic ratio of order $0.5$,
suggested that S~II is likely to contribute
no more than 2\% to the Si~II 1260~\AA\ feature.

Could the S:Si ratio in the SN1006 ejecta be substantially higher
than the cosmic ratio of $0.5$?
Recent studies of Type~Ia supernovae have favored
off-centre delayed detonation models,
in which an explosion begins as a subsonic deflagration,
which at some point develops into a supersonic detonation
\citep*{Hoflich06}.
Si and S are synthesized in the detonation phase
by partial burning to nuclear statistical equilibrium.
The overall S:Si mass ratio is $\approx 0.5$,
but the ratio can be higher at lower expansion velocities
where the burning is more complete.
For example, the S:Si ratio approaches $\sim 2$
at an expansion velocity of $6000 \  \kms$
in model CS15DD1 of \citet{Iwamoto99}.
Even at such enhanced abundance,
S would contribute less than 10\% to the Si~II 1260~\AA\ feature.

Since the chief anomaly of the Si~II 1527~\AA\ feature
is that it is missing absorption to the red side,
an explanation in terms of S~II absorption would be more plausible
if S~II were mainly unshocked, with little shocked S~II.
In that case S~II 1250.6, 1253.8, 1259.5~\AA\ would
contribute absorption to the unshocked component
on the red side of the strong Si~II 1260~\AA\ feature,
but not to the shocked Gaussian component on its blue side.
However,
the collisional ionization rate of S~II is similar to that of Si~II
at any electron temperature exceeding $\sim 3 \times 10^5 \  {\rm K}$
\citep*{Lennon88,Voronov97},
so if S~II does contribute to the Si~II 1260~\AA\ feature,
then S~II should contribute to both unshocked and shocked components
in about the same amounts as Si~II.
Thus S~II, even if present,
does not account naturally for the different line profiles of
the Si~II 1527~\AA\ and Si~II 1260~\AA\ features.
A loophole in the above argument is that
if the S:Si ratio is strongly varying with radius,
from $\sim 30$ in the unshocked gas
to $\la 3$ or so in the shocked gas,
then it might be possible to arrange that
S~II contributes appreciably only to the unshocked component
on the red side of the strong Si~II 1260~\AA\ feature.
But this possibility seems contrived.

It should be remarked that
even if S~II does contribute appreciably
to the Si~II 1260~\AA\ feature,
that would not alter the principal conclusion of this paper,
that the sharp red edge of the strong Si~II 1260~\AA\ feature
indicates a shock front.

Therefore, in the absence of any good alternative,
we are forced to the conclusion
that any discrepancy between the Si~II lines must arise
from differences between the stellar spectra of the SM star
and the comparison PG\ 0839+399 star.
Certainly this is a plausible conclusion
given the clear differences between the spectra
at wavelengths unaffected by absorption from SN1006 ejecta.
It would be useful in the future to use theoretical model atmospheres
of OB subdwarfs to resolve the discrepancies
better than we have been able to do in this paper.

\section{Summary}
\label{summary}

In this paper we have reported a high resolution
($5 \  \kms$ FWHM)
STIS spectrum of the broad Si~II, III, and IV features
produced by ejecta in SN1006,
seen in absorption against the SM star
\citep{SM80}
which lies behind and near the projected centre of the remnant of SN1006.

The Si~II 1260~\AA\ feature,
the strongest absorption feature,
is observed to have an unresolved
%or barely resolved
sharp red edge at
$7026 \pm 3 \, (\mbox{relative}) \pm 10 \, (\mbox{absolute}) \  \kms$,
supporting the proposal of Paper~1
that this edge represents the location of the reverse shock
moving into Si~II ejecta.

The shock model predicts that the position of the reverse shock
should be changing at the observable rate of
$- 2.7 \pm 0.1 \  \kms$ per year.
We have attempted to measure the change between the position of the reverse
shock over the 6 year interval between the FOS and STIS data.
The measured change of
$- 35 \pm 25 \  \kms$
is consistent with the predicted change of
$- 15.8 \pm 0.7 \  \kms$,
but is also consistent with no change.
A future precise detection of the predicted change
would confirm beyond reasonable doubt
that the Si~II 1260~\AA\ feature
indeed reveals Si~II in the process of being shocked.

The pre-shock velocity
($7026 \  \kms$),
the post-shock bulk velocity
($5160 \  \kms$),
and the post-shock velocity dispersion
($1160 \  \kms$)
of the Si~II are all measured from the Si~II 1260~\AA\ profile.
Agreement between the measured velocities and the shock jump conditions
requires that almost all the shock energy goes into ions,
with little or no electron heating.
This is consistent with other observational evidence
that little electron heating occurs in the fast shocks in SN1006
\citep{LRMB96,GWRL02,VLGRK03,Vink05}.

The STIS spectrum is consistent with the analysis of Paper~1,
which was based on the lower resolution FOS spectrum of the SM star,
and the conclusions of that paper remain unchanged.

\section*{Acknowledgements}

We thank Frank Winkler for encouraging us to write up this paper
and for comments on the manuscript,
Tom Ayres for advice on the reduction of the STIS spectra,
and Pierre Chayer for advice about the spectra of OB subdwarf stars.
Support for this work was provided by NASA through grant number
GO-7349
from the Space Telescope Science Institute,
which is operated by AURA, Inc., under NASA contract NAS 5-26555.

\end{document}